\documentstyle{llncs} 
\input epsf 
\begin{document}

\title{Set K-Cover Algorithms for Energy Efficient Monitoring in Wireless Sensor Networks}

\author{Zo\"e Abrams\inst{1}, Ashish Goel\inst{2}, and Serge Plotkin\inst{1}}

\institute{Stanford University, Computer Science Department\\
\and
Stanford University, Management Science and Engineering}

\maketitle

\begin{abstract}
Wireless sensor networks (WSNs) are emerging as an effective means for environment monitoring. This paper investigates a strategy for energy efficient monitoring in WSNs that partitions the sensors into covers, and then activates the covers iteratively in a round-robin fashion. This approach takes advantage of the overlap created when many sensors monitor a single area. Our work builds upon previous work in \cite{Potkonjak}, where the model is first formulated. We have designed three approximation algorithms for a variation of the SET K-COVER problem, where the objective is to partition the sensors into covers such that the number of covers that include an area, summed over all areas, is maximized. The first algorithm is randomized and partitions the sensors, in expectation, within a fraction $1 - \frac{1}{e}$ ($\sim$.63) of the optimum. We present two other deterministic approximation algorithms. One is a distributed greedy algorithm with a $\frac{1}{2}$ approximation ratio and the other is a centralized greedy algorithm with a $1 -\frac{1}{e}$ approximation ratio. We show that it is NP-Complete to guarantee better than $\frac{15}{16}$ of the optimal coverage, indicating that all three algorithms perform well with respect to the best approximation algorithm possible. Simulations indicate that in practice, the deterministic algorithms perform far above their worst case bounds, consistently covering more than 72\% of what is covered by an optimum solution. Simulations also indicate that the increase in longevity is proportional to the amount of overlap amongst the sensors.  The algorithms are fast, easy to use, and according to simulations, significantly increase the longevity of sensor networks.  The randomized algorithm in particular seems quite practical.\end{abstract}

\section{Introduction}
We study the problem of designing an efficient and distributed algorithm that partitions the sensors in a WSN into $k$ covers such that as many areas are monitored as frequently as possible.  The problem of choosing a cover for each sensor is abstracted into a variant of the SET K-COVER problem, in which we are given a finite set $S$ of elements, corresponding to the areas to be monitored, a collection $\{S_j\}_{j = 1}^n$ of subsets of $S$, where each $S_j$ represents a sensor and contains the areas that sensor monitors from $S$, and a positive integer $k \geq 2$.  The goal is to find a partition of the subsets into $k$ covers ${c_1, ...,c_k}$ where each cover is a set of subsets, such that $\sum_{i = 1}^k |\cup_{S_j \in c_i} S_j|$ is maximized.  Informally, we are maximizing the number of times the areas are covered by the partition.

The SET K-COVER problem can be used to increase the energy efficiency of WSNs.  A single area in a WSN may be covered by multiple sensors due to the ad hoc nature of sensor placement, topological constraints, or perhaps to compensate for the short lifetime of a sensor by placing multiple sensors close together.  Therefore, in an effort to increase the longevity of the network and conserve battery power, it can be beneficial to activate groups of sensors in rounds, so that the battery life of a sensor is not wasted on areas that are already monitored by other sensors.  In addition, certain batteries last up to twice as long when used in short bursts as opposed to continuously \cite{Benini}.  Therefore, activating a sensor only once every $k$ time units can extend the lifetime of its battery. 

Previous results on this problem~\cite{Potkonjak} solve a fair version where the objective is to  maximize $k$ such that every cover contains \emph{all} the elements.  In many environments, requiring that a cover contain all the elements may be too strict.  Consider, for instance, that there is a single area that is monitored by only one sensor but all other areas are monitored by hundreds of sensors.  Except for that single area, all other areas could be covered for much longer by dividing the sensors into covers.  But in the fair version, we cannot partition the sensors at all because only one partition would be able to monitor that one area.  Therefore, we relax the requirement that every cover contain all the elements.

We explore three algorithms that solve the SET K-COVER problem: randomized, distributed greedy, and centralized greedy.  In the randomized algorithm, each sensor simply assigns itself to a cover chosen uniformly at random from the set of all possible covers.  In the distributed greedy algorithm, each sensor assigns itself, in turn, to the cover with the minimum intersection between the areas the sensor monitors and the areas monitored by the cover thus far.  The centralized greedy algorithm is similar to the distributed greedy, except that an area in the intersection is weighted based on how likely it is to be covered by some other sensor later on in the assignment process.

The performance of our three algorithms are summarized in Table \ref{f:Sum}.
One metric for the performance of our algorithms is the worst case ratio between the number of times the areas are covered, according to the algorithm's partition, and the optimum number of times the areas can be covered by any partition.  This ratio is referred to interchangeably as the performance guarantee and the approximation ratio.  Our simulations show that for high density networks, the SET K-COVER partition can simultaneously achieve high $k$ and high coverage at each time instant.  Simulation results indicate that the increase in longevity is a constant function of the density of the network.  In Table \ref{f:Sum}, $|E|$ is the number of sensor-area pairs such that the given sensor covers the given area, and $c$ is a scaling factor(perhaps dependent on other problem parameters).  The running time of an algorithm is the number of time units the sensor network needs to create the partition (within the distributed or centralized setting in which the algorithm is presented).  $|S_{\max}|$ is the cardinality of the largest subset.  There is no worst case guarantee on fairness for the distributed and centralized greedy algorithms.  However, in simulations, calculations of the area that is covered by the least number of covers, relative to the number of sensors that are capable of covering it, suggest the algorithms are fair in practice.

\begin{table}[th]
\begin{center}
{\footnotesize{
\begin{array}[t]{|c|c|c|c|c|c|} \hline 
&&&&&\\
Algorithm          & Assump- & Approximation\;& Time & Worst\;Case& Network\\ 
                   & tions   & Ratio& &Fairness& Longevity\\ 
&   & & &Guarantee& Ratio\\
&&&&&\\ \hline \hline

&&&&&\\
    Randomized         & Minimal     &1 -\frac{1}{e} & 1 & 1 - \frac{1}{e} & \frac{|E|}{2c}\\ 
                       &            &Expected        &  &Expected&\\
&&&&&\\ \hline
&&&&&\\
&&&&None&\\
Distributed \; Greedy & Few         & \frac{1}{2}   & nk|S_{\max}|& (.5\;in\;simulations) & \frac{|E|}{c}\\ 
&&&&&\\ \hline
&&&&&\\
&&&&None&\\
Centralized \; Greedy & Many        & 1 - \frac{1}{e}          & 2nk|S_{\max}|&(.5\;in\;simulations) & \frac{|E|}{c}\\ 
&&&&&\\ \hline 

\end{array}
}}
\caption{{\footnotesize{Summary of Results.
}}\label{f:Sum}}
\end{center}
\end{table}

We find, in accordance with ``No Free Lunch Theorems''~\cite{Koza} that there is a trade-off between the complexity (both in terms of running time and simplicity) and the performance guarantee.  The randomized algorithm is remarkably simple, robust, easy to use, and easy to code.  It is also fair in two respects.
\begin{enumerate}
\item In expectation, an area is covered within $1 - \frac{1}{e}$ of the maximum number of times possible.  
\item With high probability, the least covered area is covered within $\ln n$ of the maximum number of times possible
\end{enumerate}
The randomized algorithm does bears some risk since its approximation ratio is an expectation.  The distributed greedy algorithm has a deterministic approximation ratio, but the ratio is smaller than the ratio for the randomized algorithm, and both the running time and the requirements of the network are slightly higher.  Finally, the centralized greedy algorithm gives a best possible guarantee for some variants of the problem, but it may not always possible to design a distributed implementation.  

We show that it is NP-Complete to guarantee better than $\frac{15}{16}$ of the optimal coverage, indicating that all three algorithms perform well with respect to the best approximation algorithm possible.  The hardness of approximation is obtained by a reduction from the E4-SET SPLITTING problem.  

Simulations show that in practice, the algorithms perform well above the worst case bounds proved in the theoretical analysis.  Many simulations show the algorithms covering more than 99\% of the maximum possible. 

Simulations also suggest that using the sensors in rounds has the potential to significantly increase the longevity of sensor networks.  In simulation results, the energy savings using the SET K-COVER algorithm are directly proportional to the density of the network.  Significant increases in the longevity of the network are observed when the overlap between sensors is high.  In addition, there is time gained by extended battery lifetimes due to operation in short bursts.

The paper is organized as follows.  In sections II, III, and IV respectively, a randomized, distributed greedy, and centralized greedy algorithm are presented and analyzed.  Section V shows the hardness of approximation for the SET K-COVER problem.  Section VI contains the results of various simulations.  We conclude with open problems and areas of further exploration.

\section{Randomized Algorithm}
The randomized algorithm assigns each sensor to a cover chosen uniformly at random.  It requires no preprocessing and makes extremely few assumptions about the network.  Its simplicity facilitates implementation, use, and maintenance.  It is also robust to sensor failure, and can easily accommodate the entry of new sensors into the system.  In addition, the expected coverage is high, at least $1 -\frac{1}{e}$ of the best coverage possible.  This is also true per individual area, so that the expected amount an area is covered is proportional to how many sensors are capable of monitoring that area.  We can attain close to the expected performance in practice because the algorithm is simple enough that it can be run many times during the lifetime of the sensor network.  This reduces the risk that the overall performance is far from the average. 

\noindent ASSUMPTIONS:\\
1. It is assumed that all sensors have clocks with a unified start time $t_0$, so that operations can be synchronized.\\
2. Each sensor has a random number generator.\\

The following algorithm partitions the sensors into covers and is executed in parallel at each sensor starting from initialization at time $t = 0$. 

\bigskip

{\large{
\noindent\begin{array}[b]{|l|}  \hline 
Randomized\;Algorithm\;at\;Sensor\;j\\ \hline \hline
Choose\;a\;random\;number\;i \in \{1, ...,k\}; \\  
Assign\; self\; to\; cover\; c_i; \\ \hline
\end{array}
}}

\bigskip

At the end of the algorithm, sensor $j$ belongs to cover $c_i$.  During the round-robin covering of the areas, sensor $j$ will activate itself when cover $c_i$ is active.

\begin{theorem}
\label{RA}
The expected number of times elements are covered by the randomized algorithm is a $1 - \frac{1}{e}$ approximation to OPT, where OPT is the best coverage possible.
\end{theorem}

\begin{proof}
For a single area $v$, we will calculate $E[l_v]$, the expected number of covers that cover $v$ in our solution.   We use $N_v$ to denote  the number of subsets that contain $v$.  A cover will \emph{not} contain $v$ with probability $(1 - \frac{1}{k})^{N_v}$ because there are $N_v$ sets to be assigned and each has probability $\frac{1}{k}$ of being assigned to a particular cover.  The expected number of covers containing $v$ is $k - k(1 - \frac{1}{k})^{N_v}$.  So the total expected number of times areas are covered by the partition is $\sum_v E[l_v] = \sum_v (k - k(1 - \frac{1}{k})^{N_v})$.

Let $l_v^*$ be the number of times $v$ is covered in the optimum solution.  Then, $l^*_v \leq \min(k,N_v)$ because an area cannot be covered by more than $k$ covers or by more than the number of subsets containing it.  The expected number of times areas are covered by the algorithm is at least $\sum_v E[l_v]$ and the total covered by OPT is at most $\sum_v \min(k,N_v)$.  To show the overall fraction $\frac{\sum_v E[l_v]}{\sum_v \min(k,N_v)} \geq (1 - \frac{1}{e})$ we will show that $\forall v$, $\frac{E[l_v]}{\min(k,N_v)} \geq (1 - \frac{1}{e})$.  There are two cases.

\begin{enumerate}
\item[I:] $k \leq N_v$ Then, 
$\frac{E[l_v]}{\min(k,N_v)} = 1 - (1-\frac{1}{k})^{N_v} 
\geq 1 - (1-\frac{1}{k})^k 
\geq 1 - \frac{1}{e}$.  The last inequality is due to the power series expansion of $e^x$, which shows that $(1-\frac{1}{k})^k \leq \frac{1}{e}$.

\item[II:] $k > N_v$  \\
We will show that the derivative of the ratio $\frac{E[l_v]}{\min(k,N_v)}$ with respect to $N_v$ is negative, implying that the ratio is smallest when $k = N_v$.

$\frac{d}{dN_v} (\frac{k - k(1- \frac{1}{k})^{N_v}}{N_v}) = \frac{k[(1 - \frac{1}{k})^{N_v} (1 - \ln(1-\frac{1}{k})^{N_v}) - 1]}{N^2_v}$.

This is negative iff 
$$
(1 + \ln(1-\frac{1}{k})^{-N_v}) < (1 - \frac{1}{k})^{-N_v}, \nonumber
$$
which is again true due to the power series expansion ($1 + t < e^t,t \not= 0$~\cite{Motwani}).
\end{enumerate}

\end{proof}

Another attractive property of the randomized algorithm is that the element that is 
covered least is not covered too much less than the maximum number of times that it 
could possibly be covered. From case I above, in expectation, an area is covered within $1 - \frac{1}{e}$ of the maximum number of times possible. The tails of the distribution over $l$ can also be bounded.  More precisely, consider our objective is to find a partition of the subsets of $S$ into $k$ covers such that $l$ is maximized, where $l$ satisfies $\forall v \in S, l \leq \sum_{j: v \in c_j} 1$. Let $l^*$ be the optimum value of $l$. 

\begin{lemma}
With high probability (greater than $1 - \frac{1}{n}$), the randomized algorithm gives a solution with $l \geq \frac{l^*}{24 \ln n }$.
\end{lemma}

\begin{proof}
Let $l_v$ be the number of covers area $v$ belongs to after the randomized rounding and $N_v$ be the number of sets containing $v$. $\mu_v = E[l_v]$ \\
$\geq (1 - \frac{1}{e})\min(k, N_v) \geq (1 - \frac{1}{e})l^*_v$.  Each $v$ falls into one of two cases.

\begin{enumerate}
\item[I:] $\mu_v \leq 16 \ln n$.  Then $l_v \geq 1 \geq \frac{\mu_v}{16 \ln n} \geq \frac{l^*}{\frac{e}{e-1} 16 \ln n} \geq \frac{l^*}{24 \ln n }$.
\item[II:] $\mu_v > 16 \ln n$.  Using Chernoff bounds, 
$$Pr(l_v \leq \frac{\mu_v}{2}) < exp(\frac{-\mu_v}{8}) \leq \frac{1}{n^2}$$  Because $\frac{l^*}{24 \ln n} \leq \frac{\mu_v}{2}$, $Pr(l_v \leq \frac{l^*}{24 \ln n}) < \frac{1}{n^2}$
\end{enumerate}
 
The probability that a single $l_v$ is less than $\frac{l^*}{24 \ln n}$ is less than $\frac{1}{n^2}$, so the probability that any $l_v$ is less than $\frac{l^*}{24 \ln n}$ is less than $\sum_v \frac{1}{n^2} \leq \frac{1}{n}$ due to the Bool-Bonferroni Inequalities~\cite{Motwani}. Therefore, the probability that the result does not have all $l_v$ within constant is significantly small, less than $\frac{1}{n}$.
\end{proof}

\section{Distributed Greedy Algorithm}
The distributed greedy algorithm, in contrast with the randomized algorithm, gives a \emph{deterministic} guarantee that the produced partition covers at least half as many areas as the best possible partition.  The algorithm makes some assumptions about what the network is able to do and also requires some preprocessing steps.  

\noindent ASSUMPTIONS:\\
1. A clock with a unified start time $t_0$, so that operations can be synchronized.\\
2. A unique ID number taken from the set of integers $j \in \{1, ...n\}$.\\
3. Knowledge of the parameter $k$ and memory for storing a matrix of size $k \times |S_j|$, all entries initialized to 1.\\
4. Some way to recognize an area of interest (for instance, geographic coordinates or a mapping from unique sensor information to an area identification number). \\
5. Some way to communicate with other sensors that cover a common area (preferably in a local manner through direct broadcasting).\\

\subsection{PREPROCESSING PHASE}
Several preprocessing steps must take place before the partition can be created.

First, each sensor determines which areas of interest it will be capable of monitoring once it is in an activated 'on' state.  This can be done using GPS or sensor localization which is itself an area of active research, and algorithms to achieve this task are described in ~\cite{Kleinberg},~\cite{Howard}, and~\cite{Savarese}, among others.   

Next, each sensor must determine a method of communication with other sensors covering the areas that it covers, which we will refer to as the sensor's neighbors.  It may be necessary to communicate this information using a broadcasting tree~\cite{Ashish} or other forms of message routing.  We will give a two step distributed algorithm for stationary sensors in Euclidean space with no obstacles.  However, the specific implementation of this task will vary between applications.
\begin{enumerate}
\item[1:]  Every sensor broadcasts its unique sensor ID number, the areas it monitors, and the distances to these areas, to twice the distance of the furthest area that it monitors.\\
\item[2:]  Based on information a sensor receives in step 1, from the set of sensors with which it knows it shares an area in common, it records the distance from the area to the sensor that is furthest away as its $d_j$ parameter.  If this distance is less than the distance of its broadcast in step 1, it instead sets its $d_j$ parameter to the distance that was used for broadcasting in step 1.
\end{enumerate}

This process ensures that every sensor node knows the broadcast distance necessary so that the other nodes covering a common area can be notified by the sensor.  The $d_j$ distance will be used by the sensor to inform other sensors of its decisions during the partition phase.

\subsection{PARTITION PHASE}
In this phase, the sensors are partitioned into covers. The algorithm is initiated at time $t=0$. 

\bigskip

{\small{
\begin{array}[t]{|l|} \hline 
Distributed\; Greedy\; Algorithm\; at\; Sensor\; j\\ \hline \hline
While\;t\;<\;j \\
\;\;\;If\;message\;is\;received\;that\;an\;area\;v \in S_j\;will\;be\\
\;\;\;monitored\;by\;another\;sensor\;in\;cover\;c_i,\\
\;\;\;then\;change\;the\;entry\;in\;row\;i,\;column\;v,\;\\
\;\;\;from\;1\;to\;0;\\
If\;t\;=\;j \\
\;\;\;Choose\;i \in \{1, ...k\}\;such \;that\;the \;sum \;along \;row \\
\;\;\;i\; is\; largest;\\
\;\;\;Assign\; self\; to\; cover\; c_i; \\
\;\;\;Broadcast\; information\; about\; this\; decision\; \\
\;\;\;to\; neighbors; \\ \hline
\end{array}
}} 

\bigskip

The above distributed greedy algorithm is simple and requires only $nk|S_{\max}|$ time.  In addition, it is guaranteed to cover more than half of what the optimum partition is capable of covering.

\begin{theorem} The distributed greedy algorithm is a $\frac{1}{2}$ approximation for the SET K-COVER Problem.
\end{theorem}
\begin{proof}
Proof by construction.  We will iterate back through the $n$ subsets, creating a copy of $S_j$, called $S^*_j$ at its location $c^*_i$ in OPT.  The number of newly covered elements by $S^*_j$ is $\alpha (S^*_j)$ and the number of elements covered by $S_j$ at the moment it was assigned to $c_i$ at time $t = j$ will be called $\alpha (S_j)$.  Because $S_j$ was assigned to $c_i$, and because $\alpha (S^*_j)$ only decreases by the addition of more covers as we are iterating backward, $\alpha (S^*_j) \leq \alpha (S_j)$.  In addition,  $\sum_j \alpha (S^*_j) + \sum_j \alpha (S_j) \geq OPT$ since this assignment subsumes the sets assigned to their optimal positions. Combining equations, $\sum_j \alpha (S_j) \geq \frac{OPT}{2}$.
\end{proof}

\section {Centralized Greedy Algorithm}
The centralized greedy algorithm has a better approximation ratio than the distributed greedy algorithm, and this ratio is tight for some instances of the problem.  However, the communication and storage requirements for deploying this algorithm in a distributed setting are more involved than the above algorithms and may vary greatly between applications.  We do not propose this as a distributed algorithm but instead show that in a centralized setting, the performance of the randomized algorithm can be made into a deterministic guarantee.  We leave as an open problem the implementation of this algorithm in a distributed setting.  

This algorithm is the same as the distributed greedy algorithm except that each area is assigned a weight of  $(1 - \frac{1}{k})^{y_v - 1}$ where $y_v$ is the number of subsets containing area $v$, in the given time step, that have not yet been assigned to a cover.  Now, instead of summing entries in the rows of the matrix, the matrix is multiplied with a $|S_j| \times 1$ vector corresponding to the weights of the areas covered by the sensor.  The sensor is then assigned to the column which is largest in the $1 \times k$ vector resulting from the matrix multiplication.
Through this process, the algorithm chooses a cover $c_i$, for a given subset $S_j$, that maximizes the weighted sum of uncovered elements,
$\sum_{v: v \in S_j \wedge v \not \in \cup_{S_j \in c_i} S_j } (1 - \frac{1}{k})^{y_v - 1}$, instead of simply $\sum_{v: v \in S_j \wedge v \not \in \cup_{S_j \in c_i} S_j } 1$ as in the distributed greedy algorithm.  This is an intuitive algorithm in that each subset is assigned to the cover where it covers the largest possible number of uncovered elements, weighted according to how likely it is that the element will be covered in future iterations.

\bigskip

\begin{array}[t]{|l|} \hline 
Centralized\; Greedy\; Algorithm \\ \hline \hline
Initialize\;C = \{c_1 := \emptyset, ...,c_k := \emptyset\}; \\
For \;j:=1\; until \;n\; \\
\;\;\; find\;  i = argmax_i \sum_{v: v \in S_j \wedge v \not\in \cup_{S_j \in c_i} S_j } (1 - \frac{1}{k})^{y_v - 1} ;\\
\;\;\; c_i := c_i \cup S_j\;  (assign \;S_j\; to\; the\; cover\; c_i); \\ \hline
\end{array}

\bigskip

We will prove that this algorithm gives a $1 -\frac{1}{e}$ approximation ratio by showing that the above greedy algorithm is the derandomization of random assignment using the method of conditional expectation. 

\begin{theorem} The centralized greedy algorithm is a $1 - \frac{1}{e}$ approximation for the SET K-COVER Problem.
\end{theorem}

\begin{proof}
We would like to show that at each decision, the conditional expectation, given that decision, is greater than the expectation before being conditioned on that decision.  Suppose we are at the step where we are assigning subset $S_j$.  We want to assign $S_j$ to a cover such that the expected number of areas covered, conditioned on having assigned to cover $c_i$, is maximized.  More precisely, if we denote by $a_{ji}$ the assignment of subset $j$ (in iteration $j$) to cover $c_i$ and by $p_a$ all subset-cover assignments from previous rounds, we want to choose $i$ that maximizes $\sum_v E[l_v|p_a \wedge a_{ji}]$.  Because we maximize at every step, by linearity of expectation, the conditional expectation cannot decrease.  Therefore, at the end of the algorithm, we have an assignment for which the objective function is at least expected initial value~\cite{Vazirani}.

The subset $S_j$ will only effect  $E[l_v | p_a \wedge a_{ji}]$ if it contains area $v$ so we will ignore vertices not in $S_j$ in our decision.  Suppose an area $v$ that is in subset $S_j$ is covered in exactly $x$ covers before the assignment of $S_j$.  Then the expected number of times $v$ will be covered is $E[l_v] = k - (k-x)(1 - \frac{1}{k})^{y_v}$.  Regardless of where $S_j$ is placed, $y_v$ will decrease by $1$.  If $v$ is newly covered in some cover, $x$ will increase by $1$, otherwise $x$ will remained unchanged.  Let us consider both scenarios:
\begin{enumerate}
\item[I:] Element $v$ is not newly covered by $S_j$ in the assignment $a_{ji}$.
Then,      
$$
E[l_v|p_a \wedge a_{ji}] = k - (k-x)(1 - \frac{1}{k})^{y_v-1}
$$
\item[II:] Element $v$ \emph{is} newly covered by $S_j$ in the assignment $a_{ji}$.
Then,\\
$$      
{\footnotesize{
\noindent\begin{array}[b]{ll}\\
E[l_v|p_a \wedge a_{ji}]&= k - (k-x-1)(1 - \frac{1}{k})^{y_v-1} \\
&= k - (k-x)(1 - \frac{1}{k})^{y_v-1} + (1 - \frac{1}{k})^{y_v-1}\\
\end{array}
}}
$$
\end{enumerate}

The component of the conditional expectation that our choice of assignment affects is whether or not an element falls into scenario I or II.  If it is in scenario II, the profit is the last term of the above equation, $(1 - \frac{1}{k})^{y_v-1}$.  So we want to maximize $\sum_{v: v \in S_j \wedge v \not \in \cup_{S_j \in c_i} S_j} (1 - \frac{1}{k})^{y_v-1}$.  This results in the above greedy algorithm.

We now have an algorithm that deterministically performs as well as the expected performance of the randomized algorithm.
\end{proof}

\section{Hardness of Approximation}
For specific cases, our algorithm is tight.  In particular, SET K-COVER is a generalization of the E4-SET SPLITTING problem, and it is NP-hard to design an approximation algorithm for E4-SET SPLITTING that performs better than our algorithm.  We will first show a weaker statement, that the general case cannot be approximated to better than $\frac{15}{16}$.  We will begin with some necessary definitions.

\begin{definition}
In the E4-SET SPLITTING problem we are given a ground set $V$ and a number of sets $R_i \subset V$ each of size exactly $4$.  Find a partition $V_1,V_2$ of $V$ to maximize the number of $i$ with both $R_i \cap V_1$ and $R_i \cap V_2$ nonempty.
\end{definition}

The hardness of approximation for E4-SET SPLITTING has been well studied, leading to the following result using PCP~\cite{Hastad}.

\begin{theorem}
It is NP-hard to distinguish between instances of Max E4-SET SPLITTING where all the sets can be split by some partition and those where any partition splits at most a fraction $\frac{7}{8} + \epsilon$ of the sets, for any $\epsilon > 0$.
\end{theorem}

We use the above definitions to show the hardness of SET K-COVER.

\begin{theorem}
It is NP-Complete to $\alpha$-approximate the SET K-COVER problem with $\alpha \geq \frac{15}{16} + \epsilon$ for any $\epsilon > 0$. 
\end{theorem}

\begin{proof}
Given an approximation algorithm $A$ for the SET K-COVER problem, we could use it to approximate E4-SET SPLITTING.  Suppose we would like to approximate an instance $I$ of the E4-SET SPLITTING problem.  We can create an instance $I'$ of the SET 2-COVER problem.  For every variable of the ground set $V$ in $I$, there is a subset in $I'$.  For every set $R_i \subset V$ in $I$, there is an element in the set $S$ of $I'$.  A subset in problem $I'$ contains an element of $S$ iff the corresponding variable from $V$ belonged to the corresponding set $R_i$.  
The proof is by contradiction.  Assume $\alpha = \frac{15}{16} + \epsilon$ for some $\epsilon > 0$.

Case 1:  All the sets can be split in $I$.
Then the optimum in $I'$ is $2|S|$ and we run algorithm $A$ on $I'$ and are guaranteed to cover at least $(\frac{15}{16}+ \epsilon)2|S| = (\frac{15}{8} + 2\epsilon)|S|$ elements.  

Case 2:  Only a fraction $\frac{7}{8} + \epsilon$ of the sets can be split in $I$.
Then the optimum in $I'$ is less than $(\frac{15}{8} + \epsilon)|S|$, and any solution to $I'$ will be less than this value.

Therefore, we could use $A$ to distinguish between instances of $I$ that can be split completely and instances where only a fraction $\frac{7}{8} + \epsilon$ of the sets can be split, which would contradict $Theorem\;4$.
\end{proof}

In fact, after more precise analysis of the centralized greedy algorithm in the context of E4-SET SPLITTING, we see that the algorithm achieves an approximation ratio of exactly $\frac{15}{16}$ and is therefore tight.

\begin{theorem}
The centralized greedy algorithm is the best approximation algorithm possible for specific instances of the SET K-COVER problem.
\end{theorem}

\begin{proof}
Consider instances where the number of covers is $k=2$ and every area is contained in exactly $4$ subsets, implying $N_v = 4,$ $\forall v$.  From the proof of $Theorem\;1$, the approximation ratio $\forall v$ is $\frac{E[l_v]}{\min(k,N_v)} = \frac{(k - k(1 - \frac{1}{k})^{N_v})}{k} = \frac{15}{16}$. 
\end{proof}

 The centralized greedy algorithm is therefore the best approximation possible when we constrain the parameters $k$ and $N_v$.

\section{Simulation Results}

We performed simulations using all three algorithms.  Problem instances were generated by setting parameters $|S|$ (number of areas), $n$ (number of subsets), and $|E|$ (number of edges).  Then a bipartite graph is created, where the edges are chosen uniformly at random from all possible subset-area pairs.  A subset is then considered to contain an area if it has an edge connecting it with that area. For each set of parameters, ten problem instances were generated and the numbers in the tables below are the average result over all ten instances. 

We chose this approach as opposed to an approach where areas are points in Euclidean space and sensors sense within a radius of their location (as in ~\cite{Potkonjak}) because the latter limits the variety of applications.  For instance, consider the sensors are embedded in vehicles, animals, or robots that are moving around in some physical space, then the set of problem instances are much richer and our test scenarios capture this richness of possible applications.

\subsection{Performance Compared to the Optimum}
Simulations show that in practice, when compared to the optimum, our algorithms perform better than their worst case bounds.  We bound the optimum by noting that the objective function of the optimum partition cannot be larger than $k*|S|$, since we can cover at most all the areas in all covers.  We can also not hope to achieve more coverage than there are edges.  Thus we have two possible upper bounds for the optimum objective function that are listed in the column labeled OPT bound.

\begin{table}
\begin{center}
\begin{tabular}[th]{|l|l|l||l|l|l|} \hline 
$\;\;\;$n$\;\;\;$&$\;\;\;|E|\;\;\;$&$\;$OPT bound$\;$&$\;$Random$\;$&$\;$Distributed$\;$&$\;$Centralized$\;$\\ 
&&&&$\;\;\;\;\;$Greedy&$\;\;\;\;\;$Greedy \\ \hline \hline
1000&5000 &5000&3950&4837&4832\\ \hline
1000&10000&10000&6330&7625&7647\\ \hline
1000&20000&10000&8655&9677&9727\\ \hline \hline

500 &5000&5000&3951&4626&4628  \\ \hline
500 &10000&10000&6305&7277&7296 \\ \hline
500 &20000&10000&8640&9443&9470 \\ \hline \hline

2000&5000&5000&3961&4953&4954 \\ \hline
2000&10000&10000&6345&8047&8068\\ \hline
2000&20000&10000&8665&9908&9959\\ \hline
\end{tabular}
\caption{\footnotesize{For these simulations, $|S| = 1000$ and $k = 10$.}}
\end{center}
\end{table}

Simulations indicate that the deterministic greedy algorithm achieves performance that is on the order of 10-20\% better than the randomized algorithm.  The performance of the deterministic and centralized greedy solutions are strikingly close, differing by less than 1\% in every instance of the problem that was tested.  

We see the randomized algorithm is consistent with theoretical analysis, with the worst performance achieving 63\% coverage, which is quite close to the analysis of $1 - \frac{1}{e}$. 

Both deterministic algorithms perform significantly above their worst case bounds, with the lowest ratio covering more than 72\% of the maximum possible.  Many instances perform even higher, with four instances acheiving higher than 99\% of the maximum possible.   

\subsection{Increased Network Longevity}
Our simulations used the SET K-COVER algorithms to partition the sensors into $k$ covers such that when we rotate among the $k$ covers, more than $80\%$ of the areas are covered within the sliding window of k previous time steps.  Specifically, we maximize $k$ such that the total coverage is more than $.8kn$.  Since every set belongs to some cover, every area is covered at least once every $k$ time steps.  The lifetime of our solution is compared with the straightforward approach of activating all the sensors every time step until the percent covereage over the previous $k$ time steps drops below $80\%$.  We assume that all sensors have the same amount of power initially, that their energy depletes at the same rate, and that they are all capable of lasting for several time steps. Therefore, if the SET K-COVER can achieve the specified goal of $80\%$, then this signifies the lifetime of the network is more than $k-1$ times longer than the lifetime when the straightforward approach is used.  Because we only require information from $80\%$ of the nodes on average, this approach is most valuable for WSNs where it is not necessary to collect information from all the data in every time step.  In a WSN where network longevity is of primary importance this approach uses $k$ times less energy to collect the required information. 

Our simulations try several values of $k$, which is difficult to do in a distributed setting.  However, it is possible to find a good value for $k$ in advance through simulations or mathematical properties of $k$.  WSN designers can choose $k$ such that, in expectation, the solution has the desired properties.  Alternatively, running simulations in advance allows designers to make a good choice for the value of $k$ ahead of time.
 
\begin{figure}[th]
\begin{center}
\epsfverbosetrue
\epsfxsize = 280.0pt 
\epsfbox{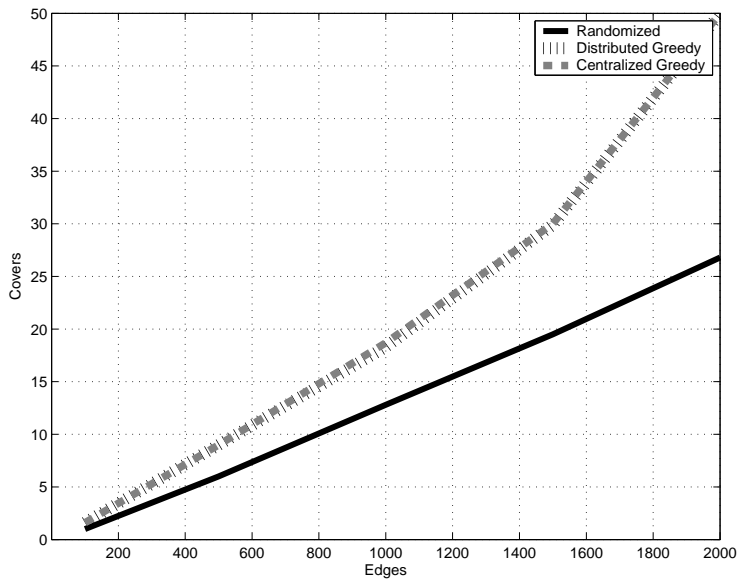}
\caption{{\footnotesize{Increase in Network Longevity.
}}\label{f:NLR}}
\end{center}
\end{figure}

Our simulations show a significant increase in the lifetime of a network that uses the SET K-COVER solution.    In Figure \ref{f:NLR}, the value of $k$ is plotted for problem instances with varying density.   For all three algorithms, the increase in longevity is proportional to the amount of connectivity.  This is expected, since a highly connected graph has more overlap and therefore more redundancy that the SET K-COVER approach can utilize to increase the lifetime of the network.  This relationship between connectivity and energy savings is reflected in the simulation results. 

\begin{figure}[th]
\begin{center}
\epsfverbosetrue
\epsfxsize = 280.0pt 
\epsfbox{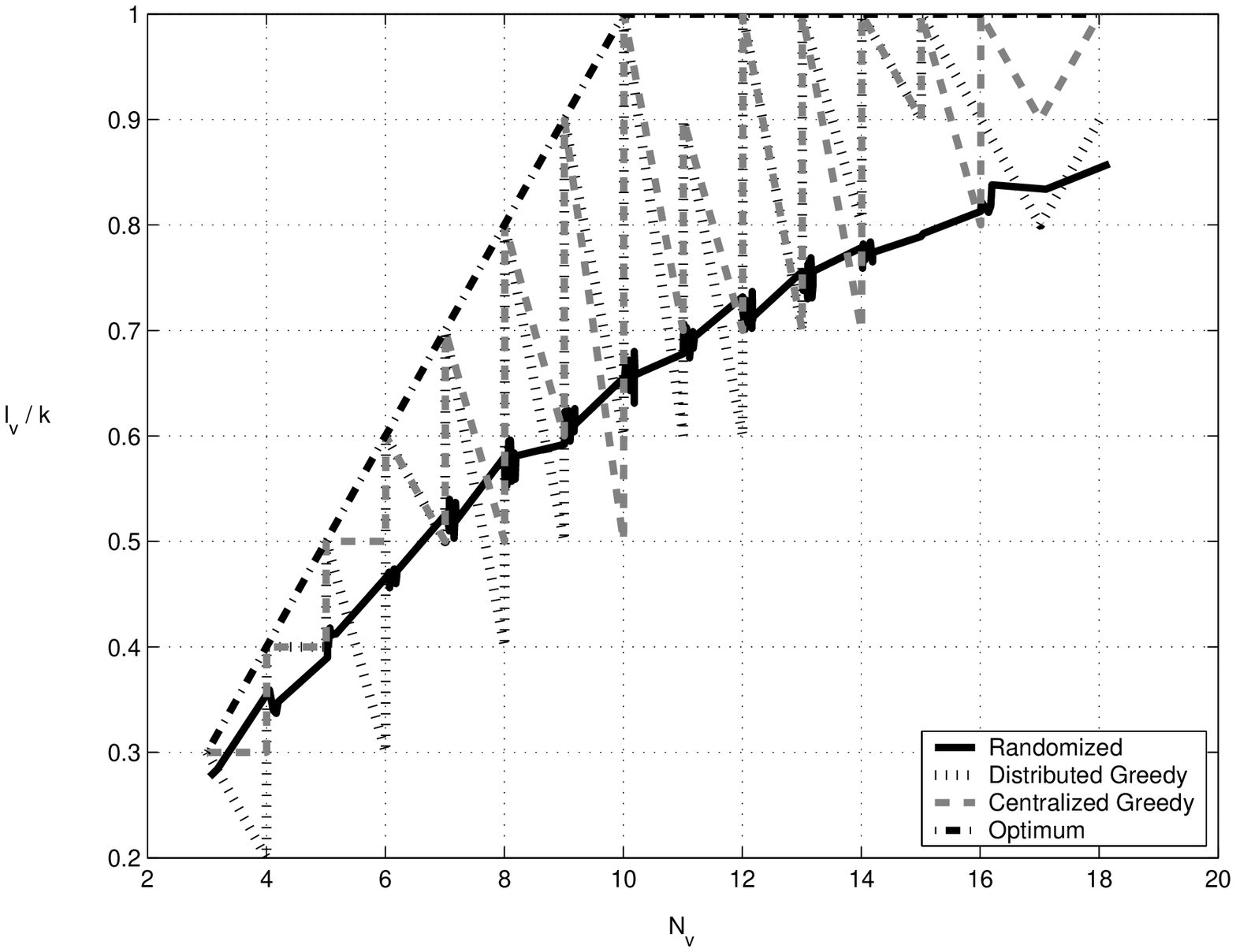}
\caption{{\footnotesize{For this problem instance, $k = 10$, $|S| = 200$,$n = 100$, and $|E| = 2000$.
}}\label{f:Fair}}
\end{center}
\end{figure}

In addition to the clear benefits in energy savings, the covers produced by our algorithms have the useful property that they result in coverage of an area that is positively correlated with the number of sensors covering that area.  This means that if there is a particular area in need of more frequent monitoring, then multiple sensors could be deployed close together to bolster the monitoring capabilities in that area.  For example, if we are monitoring traffic, we might want frequent coverage of a busy highway intersection, and have less need for vigilant sensor information about an empty country road.  Figure \ref{f:Fair} charts 200 elements for a single problem instance, with $N_v$ plotted along the domain and $\frac{l_v}{k}$ plotted along the range.  The randomized algorithm was applied 100 times on the same problem instance and the results in Figure \ref{f:Fair} are the average over all of these runs.  The optimum equals $min(\frac{N_v}{k},1)$ because an area cannot belong to more covers than the number of subsets containing it.  In the distributed and centralized greedy algorithms, no $l_v$ has a value that is less than 50\% of the optimum $l_v$ it could possibly obtain.  In the randomized algorithm, the worst ratio occurs when k = 10 and the ratio is $.63 \approx 1 - \frac{1}{e}$ in accord with theoretical analysis.  We see that on average, the $l_v$ values are within 70\% to 80\% of the optimum.   These simulations suggest that the algorithms are fair in that every area receives coverage relative to the number of sets that cover the area.

\begin{figure}[th]
\begin{center}
\epsfverbosetrue
\epsfxsize = 280.0pt 
\epsfbox{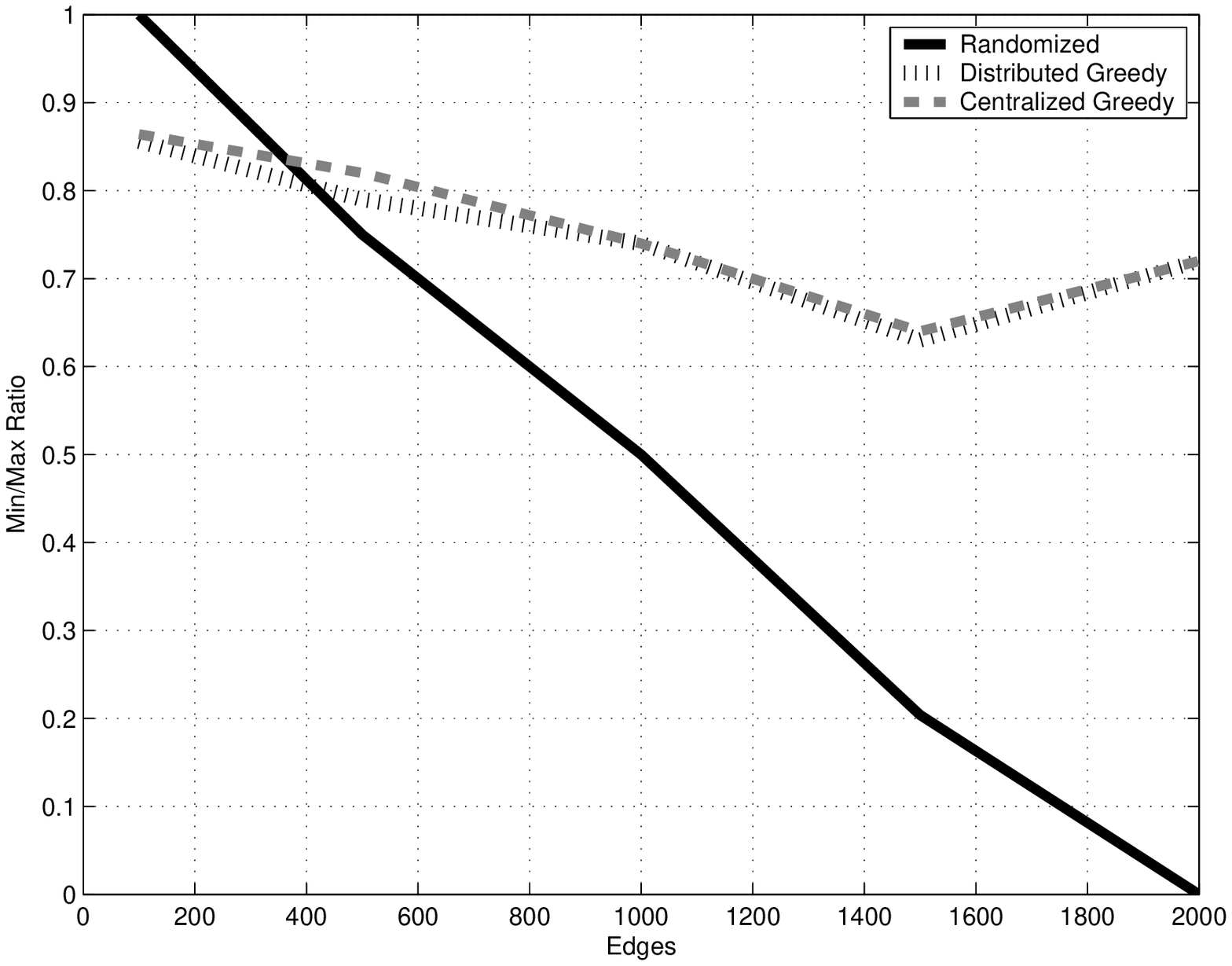}
\caption{{\footnotesize{Cover Size Ranges.
}}\label{f:CoverRange}}
\end{center}
\end{figure}

Another convenient property of the greedy algorithms is that for all covers in a given solution, the number of areas covered by each cover lies within a small range.  Thus, we could use the SET K-COVER partitions if we had the requirement that \emph{every} cover monitor at least 80\% of the areas.  In Figure \ref{f:CoverRange} we graph the size of the minimum cover divided by the size of the maximum cover, over several problem parameters.  We see that for the distributed and centralized greedy algorithms, the smallest cover is always at least 60\% of the largest cover.  When there are many covers (as in the problems with $|E| = 2000$), the ratio decreases slightly since it is more likely to have outliers when the group is larger.  When there are many covers in the randomized algorithm, however, there are a few covers with no areas at all.  As the number of covers increases, the probability there will be a cover with little or no areas becomes larger, leading to the fast dropoff we observe in Figure \ref{f:CoverRange}.  Therefore, the distributed and centralized greedy algorithms are better suited for applications that require covers that lie within a close range of coverage.

\section{Open Problems}
It is an interesting area of further research to determine whether the centralized greedy algorithm can be efficiently implemented in a distributed fashion.  The main challenge in making this algorithm distributed is that it is not clear where the $y_v$ values that determine the weights should be stored and how their values are to be updated in every round.  One possible solution is to run the preprocessing phase between every sensor assignment, but this significantly increases the communication overhead.  

From a theoretical perspective, this work raises the question of whether the centralized greedy algorithm is tight for the general case when $N_v$ are non-uniform and $k>2$.  Perhaps the recent breakthroughs in lower bound results using PCP~\cite{Guru}~\cite{Hastad} can be applied to the SET K-COVER problem.

Another open area of further study is the design of approximation algorithms for fair versions of the problem.  The approach in~\cite{Potkonjak} is to design an algorithm that maximizes $k$, such that all areas are included in every cover.  We examined a flipped variant of the problem in section II, where, given a value of $k$, the fewest number of times any element is covered is maximized.  In the optimum, the second problem subsumes the first since, by doing a binary search on $k$ and choosing the largest $k$ for which $k = l$, we have found the solution to the first problem.  However, in a distributed sensor network environment, it is very difficult to try many possible values of $k$.  More work needs to be done to give a deeper understanding of the implications of using either method.

Finally, it would be interesting to explore how to place sensors in a way that works well in conjunction with round-robin covering.

\end{document}